\documentstyle[aps,prb,psfig,twocolumn,floats]{revtex}
\begin{document}
\draft

\title{s-s*-d-wave superconductor on a square lattice and its BCS phase 
       diagram}
\author{J. Ferrer, M. A. Gonz\'alez-Alvarez}
\address{
Departamento de F\'{\i}sica, Facultad de Ciencias, Universidad de Oviedo, 
E-33007 Oviedo, Spain}
\author{J. S\'anchez-Ca\~nizares}
\address{
Departamento de F\'{\i}sica Te\'orica de la Materia Condensada, C-V, 
Universidad Aut\'onoma de Madrid, E-28049 Madrid, Spain} 

\date{\today}

\address{
\begin{minipage}[t]{6.0in}
\begin{abstract}
We study an extended Hubbard model with on-site repulsion and nearest
neighbors attraction which tries to mimic some of the experimental features of 
doped cuprates in the superconducting state. We draw and discuss the phase 
diagram as a function of the effective interactions among electrons for a wide 
range of doping concentrations. We locate the region which is 
relevant for the cuprates setting some constraints on the parameters which may 
be used in this kind of effective models. We also study the effects of 
temperature and orthorrombicity on the symmetry and magnitude of the gap 
function, and map the model onto a simpler linearized Hamiltonian, which 
produces similar phase diagrams.  
\end{abstract}
\end{minipage}}

\maketitle
\date{\today}

Accurate results using a variety of experimental techniques like, for instance, 
ARPES,\cite{ding96a,harris97a} Josephson tunneling, \cite{annett96a}
penetration depth,\cite{panagopoulos97a} or thermal conductivity 
measurements\cite{aubin97a} provide evidence on the d-wave symmetry of both the 
gap and order parameter functions of optimally doped and underdoped cuprates. 
Penetration depth measurements in slightly underdoped samples \cite{kamal94a}
determined that their critical behavior fall on the 3D XY universality class
(see also Refs. \onlinecite{uemura97a,emery95a}). This fact, which is not 
consistent with BCS weak coupling theory, indicates that the phase transition 
corresponds to the Bose-Einstein condensation of a single, complex order 
parameter field. The existence of a pseudogap with d-wave symmetry above the 
superconducting state for underdoped cuprates 
\cite{loeser96a,ding96b,puchkov96a,watanabe97a} provides 
further confirmation on the non-BCS nature of the superconducting state. 

There are several theoretical schemes which seem to fit into this 
experimental state of affairs: (a) In the magnetic scenario for the 
cuprates,\cite{scalapino95a,chubukov96a} the strong on-site repulsion among 
electrons gives rise to antiferromagnetic collective excitations. These degrees 
of freedom dress the bare interaction among the residual electrons, providing 
the pairing mechanism for the d-wave superconducting state and giving rise to 
the pseudogap in the normal state. (b) In some spin-charge separation 
theories,\cite{lee96a} the pseudogap is related to the pairing of spinons and 
the superconducting state comes about when holons condense. (c) A further line 
of thought supposes that an undetermined high-energy pairing mechanism (which 
might be dressed vertex of case (a)) gives rise to an effective low energy 
pairing Hamiltonian where the coupling constants set the d-wave superconducting 
state in an intermediate coupling region.\cite{engelbrecht97a}   

A thorough study of the phase diagram and qualitative features of a simple 
model which retain the electronic structure found in ARPES experiments is 
therefore an urgent task, if superconductivity in the cuprates has anything to 
do with scenarios (a) or (c). A BCS treatment of 
such a model should serve as a starting point for more accurate solutions which 
deal properly with the strong correlations of the intermediate coupling 
regime.\cite{engelbrecht97a} It should also be a useful tool for 
studies of transport, magnetic or optic properties of the cuprates, where bulk 
or surface impurities and other inhomogeneities must be taken into 
account.\cite{balatsky96a,cook96a,wu97a} 

We present in this article a detailed study of an effective pairing model, 
where the kinetic energy term comes from a tight-binding fit to the ARPES band 
structure of ${\rm Bi}_2{\rm Sr}_2{\rm CaCu}_2{\rm O}_{8+\delta}$ 
(Bi2212),\cite{norman95a} and we use 
either an attractive or repulsive on-site interaction plus a nearest-neighbors 
attraction term. In other words: we view the ARPES band structure of Ref. 
\onlinecite{norman95a} as the low-energy pairing action of the actual 
Hamiltonian where the strong interactions among electrons, charge fluctuations, 
etc, have renormalized the kinetic energy and interaction terms giving rise to 
a set of hopping integrals, an effective on-site repulsion and a 
nearest-neighbors attraction, while other possible terms are irrelevant. The 
only channel which has not been dealt with in this kind of experimental 
implementation of the Renormalization Group is the superconducting one, so we 
need to perform only a conventional BCS decoupling scheme (the Hartree term 
would double-count interaction effects already taken into account). This model 
generates pure s and d, as well as mixed s+s*, s+s*+d and s+s*+id 
superconducting states in different regions of the phase diagram, where s*
denotes an extended s state. We have performed most of our calculations at zero 
temperature where it is believed that the BCS approximation provides the gross 
features of the true ground state of the system, at least for small enough 
interactions.\cite{engelbrecht97b,loktev97a}

We assume a tight-binding model on a squared lattice with a dispersion relation
which keeps hopping terms up to five nearest neighbors
\begin{equation}
\xi (\vec{k}) = \sum_{i=1}^5  t_i \eta_i(\vec{k})-\mu,
\end{equation} 
where the hopping integrals $t_i$ and functions $\eta_i(\vec{k})$ are listed in 
table I of Ref. \onlinecite{norman95a}. The resulting electronic structure does 
not possess electron-hole symmetry and has a band width of about 1.1 eV. We 
vary $\mu$ to change the occupation number $n=1-x$ where $x$ is the hole doping 
in the sample. The experimental evolution of the Fermi Surface (FS) with doping 
is not inconsistent with such a rigid shift of the band structure even though 
the ascription of a FS to a system with a pseudogap is a matter of 
controversy.\cite{ding97a} We have also performed the calculations keeping 
only $t_1$ and $t_2$, and obtained results quantitatively similar to those 
explained in this article.

The interaction term has an on-site potential, which can be either positive or 
negative plus a nearest neighbors attraction
\begin{equation}
V_{int}(\vec{k}\xi)= -V_0-V_2(\cos (k_x \xi)+\cos(k_y \xi)),
\end{equation}
where $\xi$ is some typical length scale. This potential resembles the 
effective pairing interaction found in the magnetic scenario of the cuprates, 
if we set $\xi$ equal to the magnetic coherence length (and slightly larger 
than the lattice spacing, see Figure 11 of Ref. \onlinecite{scalapino95a}).
  
The BCS expressions for the thermodynamic potential and gap functions are 
\begin{eqnarray}
&\Omega(T,\mu)=&M\frac{|\Delta_s|^2}{V_0}+ 
M\frac{|\Delta_{s^*}|^2+|\Delta_d|^2}{V_2}+\nonumber \\&&+
\sum_k \left[\xi_k-E_k-\frac{2}{\beta} \ln (1+e^{-\beta E_k}) \right]\nonumber 
\\ &\Delta (\vec{k})=&\Delta_s+\Delta_{s^*} F_{s^*}(\vec{k})+
\Delta_d F_d(\vec{k}),
\end{eqnarray}
where $M$ is the number of sites, $E_k=\sqrt{\xi_k^2+|\Delta_k|^2}$ is the
dispersion relation for the quasiparticles, and 
$F_{s^*,d}=\cos(k_x \xi)\pm\cos(k_y \xi)$. 

The gap is a function of the dimensionless variable $k \xi$ for this particular
choice of the interaction term, but we believe that such a radial dependence 
also holds for any potential of the form $V(\vec{k}\xi)$, even for moderate 
values of the coupling constants. In the magnetic scenario of the 
cuprates,\cite{chubukov96a} $\xi$ is the correlation length and, hence, the
furnished gap function is not commensurate with the Brillouin zone and varies 
along with $x$. Plots of $\Delta(k)$ in the $\Gamma M$ and $MY$ directions 
for some underdoped and optimally doped Bi2212 samples,\cite{ding96a,loeser96a} 
if possible, might demonstrate if this conjecture is correct. We set 
nevertheless $\xi$ equal to the lattice spacing $a$ throughout this paper.   

The saddle point equations can now be easily written because the 
pairing interactions $V_{0,2}(k-k')$ are separable:
\begin{eqnarray}
\Delta_{s}&=&\frac{V_0}{M} \sum_k \frac{\Delta_k}{2 E_k} 
\tanh(\frac{\beta E_k}{2})        \nonumber\\
\Delta_{s^*,d}&=&\frac{V_2}{M} \sum_k \frac{\Delta_k F_{s^*,d}}{2 E_k} 
\tanh(\frac{\beta E_k}{2})
\label{tb}
\end{eqnarray}

We compare the free energy per site of the different superconducting states 
$f(T,n)=\Omega(T,\mu(n))/M+\mu(n)\,n$, to ascertain their relative stability. 

\begin{figure}
\psfig{figure=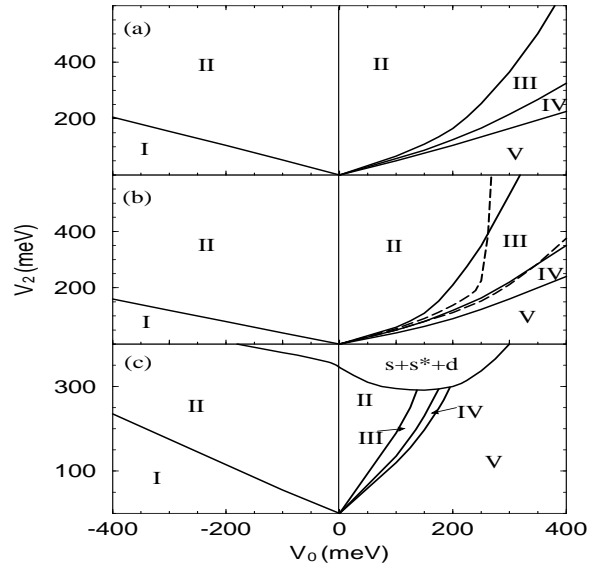,height=7cm,width=8cm}
\vspace{0.5cm}
\caption{Phase diagram for (a) $x= 0.05$ (b) $x= 0.2$ and (c) $x=0.5$. Labels
I, II and V denote the regions of metallic, d-wave and s+s* states,
while in regions III and IV both d and s+s* solutions are found.
The boundaries of the s+s*+id state are drawn only in 
figure (b), with dashed lines.}
\label{fig1a}
\end{figure}

We examine first the phase diagram $[V_0,V_d]$ for $x$ in the range [0.0-0.5] 
in order to locate the region of parameter space which can be used to best 
represent the superconducting state of the cuprates. Figure 1 shows the phase 
diagrams for x=0.05, 0.2 
and 0.5. The coupled BCS equations possess only the trivial normal state 
solution in region I; only a pure d state solution in region II; only a mixed 
s+s* solution in region V, and only a pure s solution on the positive $V_0$ 
axis. There exist both pure d and mixed s+s* solutions in regions III and IV;
the d state being more stable in region III and the mixed s+s*, in region IV. 
We also obtain a mixed s+s*+d solution for rather large values of $V_2$, and 
a mixed s+s*+id state. These two states are always the most stable as long 
as they exist. The frontiers of stability of the different solutions change 
very slightly in the doping range [0.05-0.4]. When $x$ exceeds 0.4, though, 
region II becomes increasingly thinner and localized around the $V_2$ axis. 
The only role of $V_0$ on the d-wave state is to set its domain of stability. 
In particular, the magnitude of $\Delta_d$ does not depend on it. 

The size of the gaps as a function of hole concentration for pairs $(V_0,V_2)$
inside regions II, IV and V is plotted in Figure 2. The three figures taken as 
a whole show that the d state is more stable only close to half-filling and for 
large enough values of the ratio $V_2/V_0$, while the s states exist in the 
whole range of hole dopings. The shape of the curves for $\Delta_d$ is 
independent of $V_0$; $V_2$ only changes their height, so that one can write 
$\Delta_d=f(V_2) g(x)$. They always reach a peak at $x=0.2$ and vanish at 
$x\pm 0.7$. The form of the curves for the s-state gaps does depend on both 
$V_0$ and $V_2$, on the other hand, but its two maximums always occur at 
$x=0.2$ and $0.8$. We find that the gaps of the $t_1-t_2$ model have peaks at 
the same doping concentrations. We therefore conclude that their existence and 
position is due purely to filling effects and not to the peculiarities of the 
band structure. The arrow in Fig. 2(c) marks the boundary of stability $x_c$ 
between the d and the s states, the free energy of the d state being lower for 
$x<x_c$.  

\begin{figure}
\psfig{figure=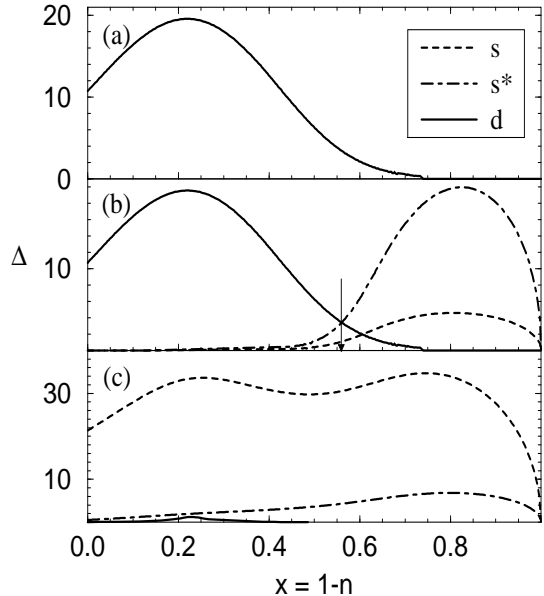,height=7cm,width=8cm}
\vspace{0.5cm}
\caption{Gaps (measured in meV) for (a) $V_0 = -250$, $V_2 = 150$ meV; 
(b) $V_0 = 50$, $V_2 = 150$ meV; and (c) $V_0 = 250$, $V_2 = 50$ meV.}
\label{fig2a}
\end{figure}

There are compelling reasons to believe that the renormalized on-site 
interaction of this low energy model should still be strongly repulsive, so we 
focus our study on the negative $V_0$ quadrant now. A first point 
to notice is that the d-wave superconducting solution disappears for values of
the on-site effective repulsion of the order of two times $V_2$. This fact 
will set some constraints on the possible parent Hamiltonian of this model
if it survives the inclusion of fluctuations over the Mean Field (MF) solution. 
Moreover, we estimate that the strength of the effective nearest neighbor 
attraction needed to obtain the experimental value of the gap for optimally 
doped Bi2212 (30 meV) is of about 140 meV, while $V_0$ can range from -320 to 0 
meV. Such values, compared to a band width of 1.1 eV, place the resulting model 
within the weak coupling regime. This result leads to question the quantitative 
accuracy of the BCS gap and of the whole MF-RPA\cite{engelbrecht97b} scheme at 
intermediate couplings even at $T=0$. What actually happens is that such 
theories provide fine estimates for the gap and other physical magnitudes for 
large values of the coupling constants while for intermediate couplings and low 
dimensions, the results are much poorer. A comparison of MF theory 
and Bethe ansatz results for the 1d Hubbard model at half filling and $T=0$ can 
be found in Ref. \onlinecite{ferrer95a}, where it is shown that the size 
of the gap is grossly overestimated by the MF solution at 
intermediate coupling.\cite{rpa} We think that the gap should be 
estimated beyond MF-RPA theory in order to make direct comparisons with 
experiments. Such a theory will require much 
larger values of $V_0$ and $V_2$ to give a gap of 30 meV, thereby placing
the coupling constants in the intermediate coupling regime.
  
The ratio $2 \Delta_d(k_{F,max},T=0)/K_B T_d$ as a function of doping is 
non-universal and larger than the value obtained for a parabolic band (4.14). 
There is a plateau, which ranges from $x=0.05$ to $0.4$, where the ratio is 
almost constant and equal to 4.35, while it grows for smaller or larger doping, 
even reaching a value of 7 for $x=0.65$. 

The present model provides a universal value $x_0=0.2$ for optimal doping,
defined as the hole concentration which yields the maximum critical 
temperature. We should caution however that experimentally, $\Delta_d(T=0)$ and 
the temperature $T^*$ at which the pseudogap begins to open up are monotonically 
decreasing functions of doping and do not peak at $x_0$, while the experimental
$T_d$ has a maximum at about $x_0=0.15$. \cite{ding97b} RPA theories
\cite{engelbrecht97b,loktev97a} of s-wave superconductors, which permit to study
the intermediate and strong coupling regimes find that electrons begin to
pair up at a temperature $T^*$ which is higher than the critical temperature
$T_c$ where pairs condense. 

\begin{table}
\caption{Changes in the gaps due to orthorrombicity for the case $V_0=-250$ 
meV, $V_2=160$ meV and $x=0.2$. Gaps are measured in meV.}
\begin{tabular}{ddddd}
$\alpha$  & $\beta$  & $\Delta_d$   &  $\Delta_s$  &   $\Delta_{s*}$  \\
\tableline
0.     &   0.     &  22.1   &   0.     &  0.    \\
0.05   &   0.     &  20.5   &  -0.28   &  0.57  \\
0.     &   0.05   &  22.1   &  -0.23   &  1.54  \\
0.05   &   0.05   &  20.6   &  -0.5    &  2.01  \\
-0.05  &   0.05   &  20.4   &   0.07   &  0.86  \\
0.05   &  -0.05   &  20.4   &  -0.07   & -0.86  \\
-0.05  &  -0.05   &  20.6   &   0.5    & -2.01  
\end{tabular}
\end{table}

We study now how orthorrombicity effects can induce mixing of d and s 
states as might happen in YBCO, a cuprate where the a and b crystal axes become
substantially inequivalent. \cite{annett96a} In order 
to do so, we redefine the 
hopping integrals $t_{i,x,y}=t_i(1\pm\alpha)$ for $i=1,3,4$ so that the 
probability amplitude for electrons to leap between two sites depends on the 
spatial direction of the hop. We also suppose that the orthorrombicity leads to 
a different nearest neighbors interaction in the $x$ and $y$ directions, 
$V_{2,x,y}=V_2(1\pm\beta)$. We find that a finite $\beta$ doesn't modify much 
the d-wave gap, but indeed induces a finite s component (see table I). 
Orthorrombicity in the hopping integrals produces appreciable changes in 
$\Delta_d$, but leads to smaller modifications in the s-state gaps. Notice 
finally how a sign change in $\alpha$ and/or $\beta$ inverts the sign of the s 
components. This effect, which was already deduced phenomenologically using 
Ginsburg-Landau theory, \cite{sigrist96a,zhitomirsky97a} implies that the 
s-wave component of the gap has opposite sign in different twin domains of 
twinned samples of YBCO. This is also the explanation  
given for the Fraunhofer pattern observed in a recent c-axis Josephson 
tunneling experiment, where the junction had been grown across a single twin 
boundary.\cite{kouznetsov97a}

One may compare the results obtained within this tight-binding model
with the
predictions of a much simpler isotropic Hamiltonian, with a linearized
dispersion relation and pairing interaction of the form
\begin{eqnarray}
&\xi_k=\frac{\hbar^2k^2}{2m}-\mu \simeq\hbar v_F (k-k_F) 
\nonumber \\
&V(\vec{k}\pm\vec{k'})\simeq -V_0^L-V_2^L\cos(2\theta)\cos(2\theta')
\end{eqnarray}
In this case, the gap function $\Delta^L=\Delta_s^L+\Delta_d^L 
\cos(2 \theta)$ obeys the following saddle point equations
\cite{musaelian95a}
\begin{eqnarray}
\Delta_{s,d}^L={\rm g}^L V_{0,2}^L \int_0^{2\pi}\frac{d\theta}{2\pi}
\int_{0}^{E_D} d\varepsilon\frac{F_{s,d}^L(\theta) \Delta^L(\theta)}
{\sqrt{\varepsilon^2+|\Delta^L(\theta)|^2}},
\label{mf}
\end{eqnarray}
where $F_s^L=1$, $F_d^L=\cos(2\theta)$, and ${\rm g}^L$ is the two-dimensional
density of states. Now, expanding Eq.\ (\ref{tb}) up to order $(k\xi)^4$ and 
comparing with Eq.\ (\ref{mf}), we obtain the following relations: 
${\rm g}^L V_0^L={\rm g}\,V_0$, ${\rm g}^L V_2^L=\gamma^2 \,{\rm g}\, V_2$, 
$\Delta_s^L=\Delta_s$ and $\Delta_d^L=\gamma\,\Delta_d$, where g is an 
effective tight-binding 
density of states and $\gamma=-(k_F\xi)^2/2+(k_F\xi)^4/24$. These two 
parameters, together with the cut-off $E_d$, can be used to fit $\Delta_{s,d}$ 
with $\Delta_{s,d}^L$ along the $V_0$ and $V_2$ axes, respectively. As shown in
Fig. 3, we indeed find linear behavior for $\log \Delta$ as a function of $1/V$, 
from which we extract values of $k_F\xi/\pi$ ranging from 1.21 to 1.27, and 
g from 1.1 to 1.6 $eV^{-1}$.

\begin{figure}
\psfig{figure=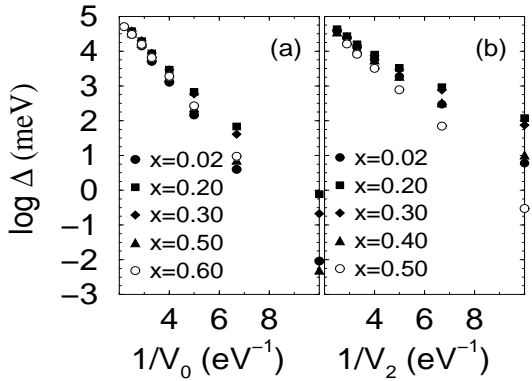,height=6cm,width=8cm}
\vspace{0.5cm}
\caption{(a) $\log \Delta_s$ versus $1/V_0$ for $V_2=0$ and several $x$. 
(b) $\log \Delta_d$ versus $1/V_2$ for $V_0=0$.}
\end{figure}

In conclusion, we have proposed and studied an effective tight-binding model 
for the superconducting state of the cuprates, guided mostly by results of 
ARPES experiments. In particular, we have delineated its weak-coupling phase 
diagrams for different hole doping concentrations close to the experimental 
optimum doping. We have been able to establish 
the region of parameters which is likely to be relevant for the cuprates in 
absolute energy units (meV). We have finally mapped the tight-binding model 
onto a much simpler linearized model and obtained similar phase diagrams.

\acknowledgments
The authors gratefully acknowledge financial support from the Spanish
Direcci\'on General de Ense\~nanza Superior, Project No. PB96-0080-C02.


\begin{thebibliography}{10}
\bibitem{rpa}
The main role of RPA is to incorporate the collective modes to the MF solution;
  RPA corrections therefore diverge in 1d even at $T=0$ because of~the
  Mermin-Wagner~theorem.
\end{thebibliography}
\end{document}